 \documentclass[showpacs, twocolumn]{revtex4}
\usepackage{bm}
\usepackage{latexsym}
\usepackage{dcolumn}
\usepackage{amsfonts,amssymb}
\usepackage{graphicx,epsfig}
\usepackage{amsmath}
\begin{document}

\def\labe{\marginpar}
\def\MA{\marginpar}
\def\a{\alpha}
\def\b{\beta}
\def\g{\gamma}
\def\d{\delta}
\def\ep{\varepsilon}
\def\vt{\vartheta}
\def\vp{\varphi}
\def\A{{\cal{A}}}
\def\ka{\kappa}
\def\pa{\partial}
\def\a{\alpha}
\def\b{\beta}
\def\m{\mu}
\def\n{\nu}
\def\g{\gamma}
\def\d{\delta}
\def\g{\gamma}
\def\la{\lambda}
\def\t{\theta}
\def\s{\sigma}
\def\r{\rho}
\def\ta{{\tilde{a}}}
\def\tb{{\tilde{b}}}
\def\tg{{\tilde{c}}}
\def\tm{{\tilde{m}}}
\def\tn{{\tilde{n}}}
\def\hi{{\hat{i}}}
\def\hj{{\hat{j}}}
 \def\hk{{\hat{k}}}
 \def\hm{{\hat{m}}}
 \def\hn{{\hat{n}}}
\def\G{{\mathcal{G}}}

\def\oG{\stackrel{{\rm o}}{\Gamma}{\!}}
 \def\oT{\stackrel{{\rm o}}{T}{\!}}
\def\LG{\stackrel{{\rm *}}{\Gamma}{\!}}
 \def\brr{\begin{eqnarray}}
\def\err{\end{eqnarray}}
\def\brn{\begin{eqnarray*}}
\def\ern{\end{eqnarray*}}


\title{No-birefringence conditions for spacetime}

\author{Yakov Itin}

\affiliation{Institute of Mathematics, Hebrew University of
  Jerusalem \\
 email: {\tt itin@math.huji.ac.il}}

\begin{abstract}
Within  the axiomatic premetric approach to classical
electrodynamics, we derive under which covariant conditions the
quartic Fresnel surface represents a unique light cone without
birefringence in vacuum.
\end{abstract}
\date{\today}
\maketitle
The classical electrodynamics theory has been reformulated
recently by  Hehl and Obukhov in an  axiomatic premetric form; for
a comprehensive account, see \cite{Birkbook} and the references
given therein. In this approach, the electromagnetic field
equations
\begin{equation}\label{feq}
dF=0\,, \qquad dH=J
\end{equation}
are accepted as consequences of the flux and charge conservation
laws respectively.
 Here $F= (1/2)F_{ij}dx^i\wedge dx^j$ is the electromagnetic strength
 2-form, while $H=(1/2)H_{ij}dx^i\wedge dx^j$ is  the electromagnetic
 excitation 2-form.  So far, the
electromagnetic field $(F,H)$ is undetermined. It has 12
independent components restricted only by 8 independent equations
(\ref{feq}). The remaining conditions are specified by the  spacetime
relation linking  the excitation to the field strength. In the
simplest case, this relation is assumed to be  local and linear,
\begin{equation}\label{rel}
H_{ij}=\frac 14 \,\kappa_{ij}{}^{kl}F_{kl}\,, \qquad H_{ij}=\frac
14 \epsilon_{ijmn}\chi^{mnkl}F_{kl}\,.
\end{equation}
 Here the tensor density $\epsilon_{ijmn}$ is $+1,-1$
or $0$ depending on whether $i,j,m,n$ is an even, an odd, or no
permutation of $0,1,2,3$. 
Recall that the physical space is considered as a bare manifold
without metric or connection. All  information on its geometry is
encoded into the constitutive tensor $\kappa_{ij}{}^{kl}$ or,
equivalently in $\chi^{ijkl}$. By definition, this tensor density 
inherits the symmetries of the electromagnetic field $(F,H)$
\begin{equation}\label{sym}
\chi^{ijkl}=\chi^{[ij]kl}=\chi^{ij[kl]}\,.
\end{equation}
Consequently, the  fourth-order constitutive tensor
$\chi^{ijkl}$ has only 36 independent components. 
Wave propagation in the premetric electrodynamics was studied by
the method of geometric optics \cite{Birkbook},
\cite{Obukhov:2000nw}. An output of this approach is a
generalized Fresnel equation
\begin{equation}\label{fres}
{\mathcal G}^{ijkl}q_iq_jq_kq_l=0
 \end{equation}
for a wave-covector $q_i$. The coefficients of this equation 
form the fourth-order Tamm-Rubilar (TR) tensor density
 \cite{Obukhov:2000nw} of weight
+1, which is completely symmetric and  cubic in the constitutive
tensor
\begin{equation}\label{tr}
{\mathcal G}^{ijkl}=\frac
1{4!}\epsilon_{mnpq}\epsilon_{rstu}\chi^{mnr(i}\chi^{j|ps|k}\chi^{l)qtu}\,.
 \end{equation}
When the premetric scheme is applied on a manifold with a
prescribed
 metric tensor $g_{ij}$, the standard Maxwell electrodynamics is
 reinstated with a special (Maxwell-Lorentz) constitutive tensor
 \begin{equation}\label{M-con}
 ^{\tt{(Max)}}\chi^{ijkl}= \lambda_0\sqrt{-g}(g^{ik}g^{jl}-g^{il}g^{jk})\,.
 \end{equation}
 Substitution of these expression in the Fresnel equation (\ref{fres})
 yields
 \begin{equation}\label{M-double}
(g_{ij}q^iq^j)^2=0\,.
\end{equation}
Consequently, for a prescribed metric, $ds^2=g_{ij}dx^idx^j$,
 the Fresnel equation yields the proper light cone which turns out
 to be a double light cone.

The idea of the premetric approach to the electrodynamics is that  the
metric tensor has to be reinstated from the pure electromagnetic data,
 in particular from the  constitutive tensor $\chi^{ijkl}$.
 An examination  of the general Fresnel equation (\ref{fres}),
 indicates that, for a  general constitutive  tensor,
 it does not  yield a unique double light cone. The birefringence
 effect of distinct light cones is known from crystal optics.
Consequently, the generalized Fresnel equation of premetric
electrodynamics predicts the possibility of birefringence in vacuum. 
Such an effect was predicted in the Lorentz-violating electrodynamics 
 \cite{Kostelecky:2002hh}, \cite{Bailey:2004na}.  When a non-zero torsion of
spacetime is coupled nonminimally to the electromagnetic field birefringence is a 
 generic effect  \cite{Preuss:2004pp},
 \cite{Rubilar:2003uf},\cite{Itin:2003hr}. 
 Moreover, classical electrodynamics modified by an axion
field, which yields a violation of Lorentz symmetry, can, if one goes
beyond the geometrical optics limit, induce birefringence of the vacuum 
 \cite{Carroll:1989vb},\cite{Itin:2004za}.

In this context, it is desirable to have the exact conditions on the
Tamm-Rubilar tensor and consequently on the constitutive tensor
which forbid the  effect of birefringence.
 In \cite{Lammerzahl:2004ww}, \cite{Hehl:2005hu}
 the absence of birefringence is
attributed by a requirement for the Eq. (\ref{fres}) to have  two
solutions of multiplicity 2. In this analysis, 
 the exact Ferrari solution of the
quartic equation was used and a {\it necessary} condition between the
coefficients was derived. We are starting from this point with an
aim to derive  a complete set of {\it necessary and sufficient}
 conditions which guarantee the uniqueness of the light cone.
Since birefringence is a proper physical effect which is
independent of a choice of a coordinate system, the
non-birefringence conditions have to be formulated in a covariant
form.

Let us decompose the wave-covector in the time and spatial
($a=1,2,3$)  components $q_i=(q_0,q_a)$. Correspondingly, the
Fresnel equation (\ref{fres}) represents a quartic surface
\begin{equation}\label{eq1}
{\mathcal
G}^{ijkl}q_iq_jq_kq_l=M_0q_0^4+M_1q_0^3+M_2q_0^2+M_3q_0+M_4=0\,,
 \end{equation}
 where
 \begin{eqnarray}\label{coef}
M_0&=&\G^{0000}\,,\qquad \quad\,\,  M_1=4\G^{000a}q_a\,,\nonumber\\
M_2&=&6\G^{00ab}q_aq_b\,,\quad
M_3=4\G^{0abc}q_aq_bq_c\,,\nonumber\\
M_4&=&\G^{abcd}q_aq_bq_cq_d\,.
\end{eqnarray}
A generic quartic polynomial with real coefficients can be always
decomposed into a product of two quadratic polynomials with real
coefficients. Let us write this decomposition as
\begin{equation}\label{eq1xx}
{\mathcal
G}^{ijkl}q_iq_jq_kq_l=M_0(q_0^2+aq_0+b)(q_0^2+cq_0+d)\,.
 \end{equation}
Since the Fresnel surface is defined only up to a scalar factor, 
 we can chose 
 \begin{equation}\label{M0}
 M_0>0
  \end{equation}
 without loss of generality. 
Due to the Fresnel equation, these quadratic factors  can vanish
independently. Since every factor represents a relation between
the components of an arbitrary wave-covector, it determines a
metric on the manifold. The quadratic factors in (\ref{eq1xx}) can
be of two types --- positive definite or indefinite.
Consequently, we consider four distinct possibilities:

(i) Both factors are indefinite and coincide. Consequently, the
induced metric is Lorentzian and unique.

(ii) Both factors are indefinite and  distinct. There are two
distinct Lorentzian metrics, i.e., there is birefringence in wave
propagation.

(iii) One factor is positive definite while the second is
indefinite.  There are two metrics --- one Lorentzian and one
Euclidean.

 (iv) Both factors are positive definite, i.e., two metrics are
 Euclidean.

 Our aim is to extract the first possibility, which corresponds to
 a unique (double) light cone.
In the cases (ii) and (iii), the sign of the left hand side of
(\ref{fres}) is indefinite. Consequently, we can remove these two 
cases by the requirement:

\vspace{0.2 cm}

\noindent {\it {\underline {Condition 1}} The Fresnel quartic
form has to be positive definite, i.e., for an arbitrary
 covector $q=q_idx^i$}
  \begin{equation}\label{cond1}
\G^{ijkl}q_iq_jq_kq_l\ge 0\,.
  \end{equation}
\vspace{0.2 cm}
Thus we come to a condition
 \begin{equation}\label{non-bier1}
 M_0q_0^4+M_1q_0^3+M_2q_0^2+M_3q_0+M_4=M_0(q_0^2+aq_0+b)^2\,.
 \end{equation}
Equating the coefficients of the same powers of $q_0$
 on both sides of (\ref{non-bier1}) we have
 \begin{eqnarray}\label{eq3a}
 &&\frac {M_1}{M_0}= 2a\,, \qquad \frac {M_2}{M_0}=a^2+2b\,,\\
 \label{eq3b}
 &&\frac {M_3}{M_0}=2ab\,, \qquad \frac {M_4}{M_0}=b^2\,.
\end{eqnarray}
 From (\ref{eq3a}),
\begin{equation}\label{eq4a}
a=\frac {M_1}{2M_0}\,, \qquad b=\frac
{4M_0M_2-M_1^2}{8M_0^2}\,.
 \end{equation}
Substituting into   (\ref{eq3b}), we derive two relations between
the coefficients of (\ref{eq1}), 
 \begin{equation}\label{eq6}
M_3=\frac {M_1}{8M_0^2} (4M_0M_2-M_1^2)
  \end{equation}
  and
 \begin{equation}\label{eq5}
M_4=\frac {(4M_0M_2-M_1^2)^2}{64M_0^3}\,.
  \end{equation}
 If we square (\ref{eq6}) and divide by (\ref{eq5})), then, for
$M_1\ne 0$, we find 
 \begin{equation}\label{eq5x}
M_4=M_0\frac{M_3^2}{M_1^2}\,.
  \end{equation}
The relation (\ref{eq6}) was derived in \cite{Lammerzahl:2004ww}
by an alternative method.

Under  the conditions (\ref{eq6}),(\ref{eq5}), the  case of two
positive definite factors is still permitted. In order to remove
this possibility, we apply an additional condition

\vspace{0.2 cm}

\noindent {\it {\underline {Condition 2}} There is a nonzero
covector $q=q_idx^i$ such that }
  \begin{equation}\label{cond2}
\G^{ijkl}q_iq_jq_kq_l= 0\,,
  \end{equation}

\vspace{0.2 cm}

Equivalently,  the roots of the
 quadratic polynomial $q_0^2+aq_0+b$ have to be real and of the
 opposite signs.
Consequently, we have an inequality
\begin{equation}\label{eq1x}
b<0,,
\end{equation}
which is equivalent to
   \begin{equation}\label{eq7}
4M_0M_2<M_1^2\,.
  \end{equation}
This condition also guarantees  that the roots of the polynomial are 
real. Indeed, if the roots are complex, they are necessary conjugate, hence 
 their product is positive. 
 Observe that for nonzero $M_1, M_3$, the inequality
(\ref{eq7}) is equivalent to
  \begin{equation}\label{eq9x}
M_1M_3<0\,.
 \end{equation}

Consequently, the relations (\ref{eq5}), (\ref{eq6}) together with
the inequality (\ref{eq9x}) guarantee uniqueness of the light
cone, i.e., the absence of the birefringence effect.

 Substituting (\ref{eq4a}) into (\ref{non-bier1}) we derive
  \begin{equation}\label{metr1}
 q^2_0+\frac{M^a}{2M}q_0q_a-\frac 1{8M^2}(M^aM^b-4MM^{ab})q_aq_b= 0\,,
  \end{equation}
  where we use the notations
  \begin{eqnarray}\label{cov}
  M&=&\G^{0000}\,,\quad M^a=4\G^{000a}\,,\quad
  M^{ab}=6\G^{00ab}\,.
  \end{eqnarray}
 Consequently, we come to a metric tensor $g^{ij}$ with the components
 \begin{eqnarray}\label{metr2}
 g^{00}&=&1\,,\qquad g^{0a}=\frac {M^a}{4M}\nonumber\\
 g^{ab}&=&-\frac 1{8M^2}(M^aM^b-4MM^{ab})\,,
 \end{eqnarray}
 which coincides with \cite{Lammerzahl:2004ww}. 
 The Lorentz nature of this metric is clear from (\ref{eq7}).
  Since the metric has been derived from the light cone
structure, it is  only determined up to a scalar factor.
 The metric tensor (\ref{metr2}) can be treated as a square root of the
 positive definite tensor $\G_{ijkl}$. Indeed, 
  \begin{equation}\label{metr3}
 \G^{ijkl}q_iq_jq_kq_l=M\left(g^{ij}q_iq_j\right)\left(g^{kl}q_kq_l\right)\,.
  \end{equation}
It is straightforward to check that this equation holds when
(\ref{metr2}) is substituted.

  Let us look for what values of the coefficients $M_i$ the equation
(\ref{eq1})  gives a unique light cone which is symmetric under a change
of the time direction $t\to -t$. It means that the incoming (past) light cone
 and the outgoing (future) light cone have the same angle.
 For this, (\ref{eq1}) has to
have two real solutions  of the same absolute value and of
opposite signs, i.e. the
 parameter $a=0$. From (\ref{eq4a}), we have
  \begin{equation}\label{t1}
 M_1=M_3=0\,\qquad M_4=\frac{M_2^2}{4M_0}\,.
 \end{equation}
 The additional condition (\ref{eq7}) takes the form
   \begin{equation}\label{t2}
M_0M_2<0\,,
  \end{equation}
 The metric tensor components (\ref{metr2}) are simplified to
 \begin{eqnarray}\label{t3}
 g^{00}&=&1\,,\qquad g^{0a}=0\,,\qquad
 g^{ab}=\frac {M^{ab}}{2M}\,.
 \end{eqnarray}
 Since the light cone is defined only up to a scalar factor,
 the time symmetric light element can be taken as
  \begin{equation}\label{t4}
ds^2=M_0dt^2-\frac 12 \,M_{ab}\,dx^adx^b\,,
  \end{equation}
  where the matrix $M_{ab}$ is the inverse of $M^{ab}$.

 As a result we have derived two covariant conditions on the Fresnel surface 
 to represent a unique light cone without birefringence.
\section*{Acknowledgment}
 I would like to thank Friedrich Hehl for most fruitful discussions. 


\begin{thebibliography}{99}
\bibitem{Birkbook} F.W.~Hehl and Yu.N.~Obukhov, {\it
    Foundations of Classical Electrodynamics: Charge, Flux, and
    Metric} (Birkh\"auser: Boston, MA, 2003).

\bibitem{Obukhov:2000nw}
  Y.~N.~Obukhov, T.~Fukui and G.~F.~Rubilar,
  Phys.\ Rev.\ D {\bf 62}, 044050 (2000)
  [arXiv:gr-qc/0005018].

\bibitem{Kostelecky:2002hh}
  V.~A.~Kostelecky and M.~Mewes,
  Phys.\ Rev.\ D {\bf 66}, 056005 (2002)
  [arXiv:hep-ph/0205211].

\bibitem{Bailey:2004na}
  Q.~G.~Bailey and V.~A.~Kostelecky,
  Phys.\ Rev.\ D {\bf 70}, 076006 (2004)
  [arXiv:hep-ph/0407252].
 
\bibitem{Preuss:2004pp}
  O.~Preuss, M.~P.~Haugan, S.~K.~Solanki and S.~Jordan,
  Phys.\ Rev.\ D {\bf 70}, 067101 (2004)
  [arXiv:gr-qc/0405068].

\bibitem{Rubilar:2003uf}
  G.~F.~Rubilar, Y.~N.~Obukhov and F.~W.~Hehl,
  Class.\ Quant.\ Grav.\  {\bf 20}, L185 (2003)
  [arXiv:gr-qc/0305049].

\bibitem{Itin:2003hr}
  Y.~Itin and F.~W.~Hehl,
  Phys.\ Rev.\ D {\bf 68}, 127701 (2003)
  [arXiv:gr-qc/0307063].

\bibitem{Carroll:1989vb}
  S.~M.~Carroll, G.~B.~Field and R.~Jackiw,
  Phys.\ Rev.\ D {\bf 41}, 1231 (1990).
\bibitem{Itin:2004za}
  Y.~Itin,
  Phys.\ Rev.\ D {\bf 70}, 025012 (2004)
  [arXiv:hep-th/0403023].

\bibitem{Lammerzahl:2004ww}
  C.~Lammerzahl and F.~W.~Hehl,
  Phys.\ Rev.\ D {\bf 70}, 105022 (2004)
  [arXiv:gr-qc/0409072].

\bibitem{Hehl:2005hu}
  F.~W.~Hehl and Y.~N.~Obukhov,
  arXiv:gr-qc/0508024.


\bibitem{Itin:2004qr}
  Y.~Itin and F.~W.~Hehl,
  Annals Phys.\  {\bf 312}, 60 (2004)
  [arXiv:gr-qc/0401016].

 \end{thebibliography}
\end{document}